\documentclass[10pt,conference,a4paper]{IEEEtran}

\usepackage{fancyhdr} 
\usepackage[USenglish,american]{babel}
\usepackage{epsfig,graphics,subfigure,graphicx,latexsym,longtable,amsmath,amscd,latexsym,amssymb,mathrsfs,syntonly,eucal}
\usepackage{multirow}
\usepackage[usenames]{color}
\usepackage[T1]{fontenc}
\usepackage{bm,cite}
\usepackage{amsbsy}
\usepackage{latexsym}
\usepackage{wasysym}
\usepackage{placeins}
\usepackage[lined,boxed,commentsnumbered]{algorithm2e}
\usepackage{lipsum}
\usepackage{url}
\usepackage{array}
\usepackage{tabu}
\SetKwInput{KwInput}{Input}
\SetKwInput{KwOutput}{Output}
\usepackage{booktabs,siunitx}
\usepackage{geometry}
\usepackage{listings}
\definecolor{codegreen}{rgb}{0,0.6,0}
\definecolor{codegray}{rgb}{0.5,0.5,0.5}
\definecolor{codepurple}{rgb}{0.58,0,0.82}
\definecolor{backcolour}{rgb}{0.95,0.95,0.92}

\lstdefinestyle{mystyle}{
    backgroundcolor=\color{backcolour},   
    commentstyle=\color{codegreen},
    keywordstyle=\color{magenta},
    numberstyle=\tiny\color{codegray},
    stringstyle=\color{codepurple},
    basicstyle=\ttfamily\footnotesize,
    breakatwhitespace=false,         
    breaklines=true,                 
    captionpos=b,                    
    keepspaces=true,                 
    numbers=none,                    
    numbersep=10pt,                  
    showspaces=false,                
    showstringspaces=false,
    showtabs=false,                  
    tabsize=1
}

\UseRawInputEncoding
\usepackage{xcolor}
\usepackage{soul}

\makeatletter
\newcommand\footnoteref[1]{\protected@xdef\@thefnmark{\ref{#1}}\@footnotemark}
\makeatother

\usepackage{bm,cite}
\usepackage{amsmath}
\usepackage{amsbsy}
\usepackage{latexsym}
\usepackage{amssymb}
\usepackage{wasysym}
\usepackage{mathtools}

\DeclareMathAlphabet{\mathbit}{OML}{cmr}{bx}{it}
\DeclareMathAlphabet{\mathsf}{OT1}{cmss}{m}{n}
\DeclareMathAlphabet{\mathTXf}{OT1}{cmss}{bx}{it}

\DeclareMathOperator{\Transpose}{T}

\DeclareMathOperator*{\argmax}{arg\ max}

\DeclareMathAlphabet{\mathpzc}{OT1}{pzc}{m}{it}

\newcommand{\Tr}{{\Transpose}}

\newcommand{\He}{{{\text{H}}}}

\graphicspath{{./figures/}}

\title{5G NR Positioning with OpenAirInterface: Tools and Methodologies}

\author{
\IEEEauthorblockN{Rakesh Mundlamuri\IEEEauthorrefmark{1}\IEEEauthorrefmark{2}, Rajeev Gangula\IEEEauthorrefmark{2}, Florian Kaltenberger\IEEEauthorrefmark{1}\IEEEauthorrefmark{2}, and Raymond Knopp\IEEEauthorrefmark{1} 
}
\IEEEauthorblockN{\IEEEauthorrefmark{1}Communication Systems Department,
EURECOM, Biot, France
}

\IEEEauthorblockN{\IEEEauthorrefmark{2}Institute for the Wireless Internet of Things, Northeastern University, Boston, USA 
}
}

\begin{document}
\maketitle

\begin{abstract}

The fifth-generation new radio (5G NR) technology is expected to provide precise and reliable positioning capabilities along with high data rates. The Third Generation Partnership Project (3GPP) has started introducing positioning techniques from Release-16 based on time, angle, and signal strength using reference signals. However, validating these techniques with experimental prototypes is crucial before successful real-world deployment. This work provides useful tools and implementation details that are required in performing 5G positioning experiments with OpenAirInterface (OAI). As an example use case, we present an round trip time (RTT) estimation test-bed based on OAI and discusses the real-word experiment and measurement process.

\end{abstract} 

\section{Introduction} \label{sec:intro}

Besides offering high data rate communication services, 
the fifth-generation new radio (5G NR) and beyond technologies
are expected to offer precise positioning and sensing capabilities. 
The third generation partnership project (3GPP) has begun defining 5G NR positioning methods and target accuracies for various scenarios starting from Release-16\cite{3gpp2018nr_38_855,3gpp2018nr_38_455_16,3gpp2018nr_38_857,3gpp2018nr_38_455_17,WanHuaYiCh_23,DwivediEtal_21}. Timing-based 5G position methods include Enhanced cell ID (E-CID), Downlink time difference of arrival (DL-TDoA), Uplink time difference of arrival (UL-TDoA), and Multi-cell or single-cell round trip time (RTT).
Unlike TDoA methods, very tight synchronization among participating gNBs is not required in RTT-based methods.

Despite the standardization of these 5G positioning methods, their performance evaluation is relatively limited to system-level simulations or a few proprietary real-world experimental evaluations \cite{DwivediEtal_21,qualcomm1,qualcomm2}. On the other hand, open-source 5G platforms such as OpenAirInterface (OAI) \cite{KalaloAbhiLuh_20} and srsRAN \cite{srsran} are playing a crucial role in experimental research. The ability to run the 5G protocol stack on general-purpose computing platforms paired with software-defined radios (SDRs) makes them an attractive tool for researchers and prototype developers.

Few works have demonstrated the timing-based positioning techniques in real-world experiments using the OAI platform \cite{del2022proof,del2023preliminary,del2023first,ahadi20235g,9963853,LiChuWaZh_22,malik2024conceptreality5gpositioning,Mundlamuri2024}. For instance, the works \cite{del2022proof,del2023preliminary,del2023first} demonstrated DL-TDOA-based positioning using positioning reference signal (PRS), and the works \cite{ahadi20235g,malik2024conceptreality5gpositioning} have demonstrated UL-TDoA-based positioning using sounding reference signal (SRS). E-CID-based positioning using a random access channel has also been demonstrated in \cite{LiChuWaZh_22}. Finally, a new RTT scheme has been proposed in \cite{Mundlamuri2024}. This method is shown to be robust to clock drift and has reduced latency in estimating the RTT compared to the existing schemes in the standards.

All these works \cite{del2022proof,del2023preliminary,del2023first,ahadi20235g,9963853,LiChuWaZh_22,Mundlamuri2024}, however, lack a comprehensive documentation on the implementation details of the positioning methods within the OAI framework. The authors in \cite{malik2024conceptreality5gpositioning} have attempted to address this issue by providing a detailed explanation of the positioning procedures implemented at the protocol level in OAI and a tutorial for performing positioning within the framework. However, detailed physical layer implementation aspects, such as fixed-point representation and device calibration, which are crucial for prototyping positioning schemes, are missing. For a researcher trying to develop positioning prototypes, navigating through a vast open-source codebase like OAI, understanding the reference signal implementation, calculating metrics such as signal-to-noise ratio (SNR), and calibrating the hardware might be daunting. Furthermore, tools to extract required data from OAI without affecting the system's performance are crucial in prototype development.

In this work, we provide a comprehensive guide on the physical layer implementation aspects and present the key functions related to the positioning in OAI. Building on these tools, we present the positioning-based prototype development and experimentation process in OAI through an example of our recent work in RTT estimation \cite{Mundlamuri2024}. We discuss the data extraction from OAI during the experiment and present a dataset collected from the testbed. Specifically, our contributions are as follows:
\begin{itemize}
    \item Present the essential functions specific to the reference signals relevant for positioning in OAI.
    \item Describe the physical layer implementation aspects in OAI.
    \item Present the {\tt{T\_tracer}} tool in OAI to extract the required data. 
    \item Present a prototype based on the RTT and discuss the dataset collected during the experiment with this prototype.
\end{itemize} 

The rest of the paper is organized as follows: Section~\ref{sec:ref_sginals} reviews the reference signals relevant to 5G positioning in OAI. In Section~\ref{sec:phy_layer_analysis}, we describe the physical layer implementation aspects in OAI. Section~\ref{sec:data_extraction} introduces a data extraction tool in OAI. Section~\ref{sec:prototype} outlines the experimental setup of a RTT-based positioning prototype and the datasets collected using the prototype are discussed in Section~\ref{sec:dataset}. Finally, Section~\ref{sec:conclusion} concludes the paper.

\begin{figure*}
    \centering    \includegraphics[width=\textwidth,]{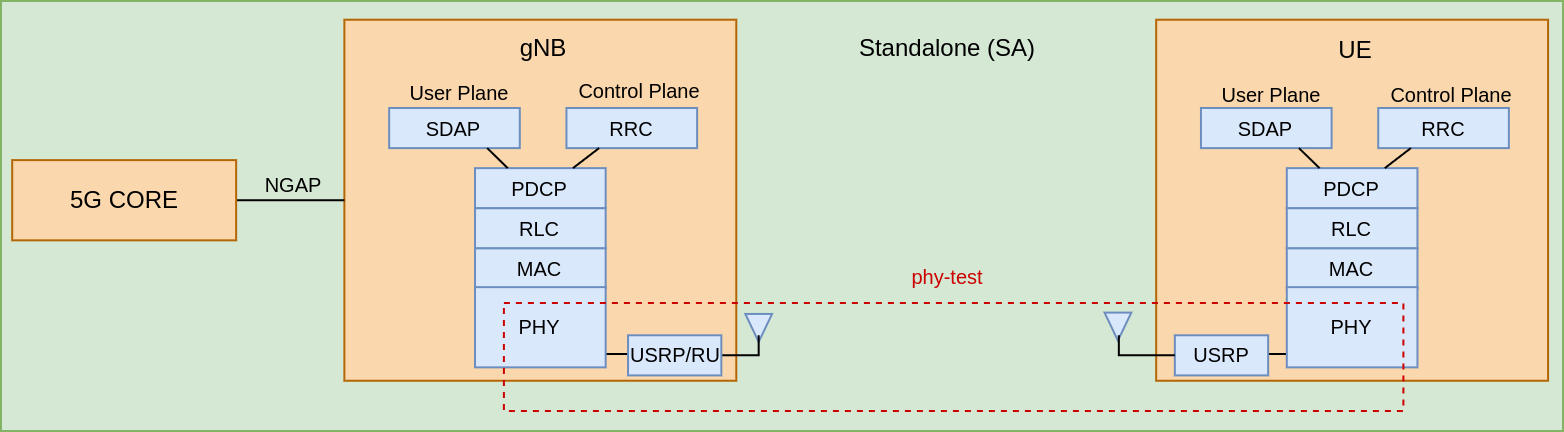}
    \caption{OAI Components.}
    \label{fig:oai_comp}
\end{figure*}

\begin{figure*}
    \centering    \includegraphics[width=\textwidth,]{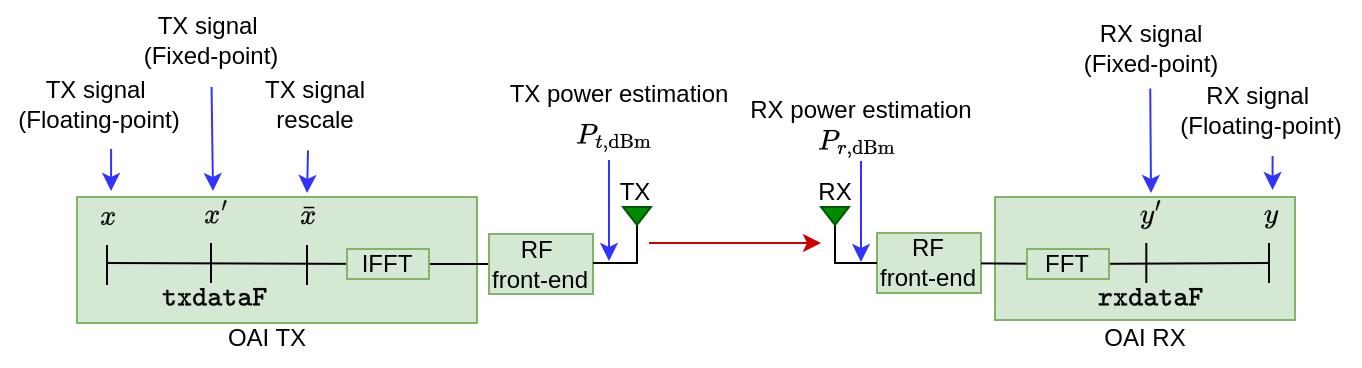}
    \caption{OAI Transmission and Reception.}
    \label{fig:link_budget}
\end{figure*}

\section{Reference Signals}\label{sec:ref_sginals}
This section provides an overview of the various components of OpenAirInterface (OAI) software, its operating modes, and
reference signals that are relevant to 5G positioning. 

\subsection{OAI Components and modes}

An end-to-end 5G network can be setup with OAI using its software components: Core network (CN), base station (gNB), and user equipment (nrUE).
To support the development, debugging and real-time experimentation process, OAI supports several modes of operation. Specifically, the following two modes are widely used in over-the-air experiments and are illustrated in Figure~\ref{fig:oai_comp}.
\begin{itemize}
    \item The {\tt{sa}} mode enables the establishment of an end-to-end 5G network in standalone mode. Either a nrUE or a commercial-off-the-shelf (COTS) UE can connect to the OAI 5G network.   
    \item The {\tt{phy-test}} mode, on the other hand, is designed specifically for testing the physical layer of the gNB and nrUE by abstracting the higher layers. 
\end{itemize}

Using the {\tt{phy-test}}, the developers and researchers can test and validate physical layer implementations and algorithms without worrying about the higher layer procedures.
The upper layer abstraction is achieved by sharing a configuration file containing higher layer parameters between the gNB and nrUE. The instructions to setup these modes are described in \cite{phy_test}.

Further, the key reference signals utilized for positioning in 5G are discussed below.

\subsection{PRACH}
The random access channel (RACH) is used for uplink synchronization.
While the UE initiates the random access (RA) procedure for initial access, the gNB can order the UE to initiate an RA procedure in the event of loss of UL synchronization.
By detecting the RACH preamble, the gNB can infer a coarse RTT between the gNB and UE.
The RACH signal is generated using the Zadoff-chu base sequence. Depending on the sequence length and repetitions, 
several formats are defined in 5G. 
The supported Formats in OAI are: 0, 1, 2, 3, A1, A2, A3, B1, B2, B3. The functions 1-4 in Table~\ref{tab:ref_sig} provide the implementation details in OAI.

\subsection{PRS}
3GPP has introduced the PRS specifically for localization purposes in the DL\cite{3gpp2018nr_38_211}. PRS is generated using QPSK modulated 31-length gold sequence and can be flexibly arranged in any number of physical resource blocks in the frequency domain. In the time domain, the PRS resources can span \{2,4,6,12\} consecutive OFDM symbols. 
However, When it comes to OAI implementation,
PRS feature is currently limited to {\tt{phy-test}} mode usage as higher layer procedures supporting PRS are yet to be implemented. Functions 5 and 6 in Table~\ref{tab:ref_sig} provide implementation details of the PRS. Detailed instructions for configuring and running the PRS can be found in \cite{prs}.



\subsection{SRS}
SRS is a wideband reference signal transmitted by the UE in the UL for channel estimation and positioning purposes. SRS is generated using the Zadoff-Chu sequence and can be flexibly arranged in the frequency domain based on a few radio resource control parameters. In the time domain, the SRS resources can span \{1,2,4\} consecutive OFDM symbols. SRS can be configured to transmit periodically or aperiodically. Currently, OAI supports periodic SRS configuration and can be operated in both {\tt{sa}} and {\tt{phy-test}} mode. Functions 7-10 in Table~\ref{tab:ref_sig} implements the SRS procedures in OAI.
\begin{table}[htbp]
\caption{Reference signals in OAI}
\centering
\begin{tabular*}{\columnwidth}{@{\extracolsep{\fill}}ll}
\toprule
Function & Description\\
\midrule
1. {\tt{nr\_ue\_prach\_procedures()}} & Initiates RACH TX procedures \\
2. {\tt{generate\_nr\_prach()}} & Generates the RACH sequence\\
3. {\tt{L1\_nr\_prach\_procedures()}} & Initiates RACH RX procedures \\
4. {\tt{rx\_nr\_prach()}} & RACH preamble detection\\
5. {\tt{nr\_generate\_prs()}} & PRS sequence generation \\
6. {\tt{nr\_prs\_channel\_estimation()}} & PRS channel, ToA estimation\\
7. {\tt{ue\_srs\_procedures\_nr()}} & Initiates SRS TX procedures \\
8. {\tt{generate\_srs\_nr()}} & Generates SRS sequence \\
9. {\tt{nr\_srs\_channel\_estimation()}} & SRS channel, ToA estimation\\
10. {\tt{configure\_periodic\_srs()}} & Configure periodic SRS\\
\bottomrule
\end{tabular*}
\label{tab:ref_sig}
\end{table}

We now describe the baseband representation of these signals in OAI.

\section{OAI Physical Layer}\label{sec:phy_layer_analysis}

Let us consider a scenario where a single antenna OAI transmitter (TX)
communicating with a single antenna receiver (RX) using 5G NR protocol stack as shown in Figure~\ref{fig:link_budget}.
This models both the uplink (UL) and downlink (DL) communication scenarios. For example, when the TX is a gNB and the RX is a nrUE it represents the DL scenario.
The received signal $y[k]$ at the RX on the $k$-th, $k\in[0,K-1]$, resource element (RE) is represented as,
\begin{equation}
    y[k] = h[k]x[k] + n[k],
\end{equation}
where, $K$ is the total number of REs, $h[k]$ represents the baseband propagation channel, $x[k]$ denotes the transmitted symbol and $n[k]$ is additive white Gaussian noise.
We now present the fixed-point format used in OAI to represent these signals. This knowledge is crucial in OAI-based positioning experiments and system design.

\subsection{Baseband Signal Representation}
The frequency domain IQ samples at the TX are stored in a contiguous memory buffer {\tt{txdataF}} as
\begin{equation}
    \text{{\tt{txdataF}}} = [\text{I}_0 \text{Q}_0\text{I}_1\text{Q}_1\dots \text{I}_{K-1}\text{Q}_{K-1}],
\end{equation}
where $\text{I}_k$ and $\text{Q}_k$ represent the in-phase and quadrature-phase components of the $k$-th RE, $k\in[0,K-1]$. Each $\text{I}_k$ and $\text{Q}_k$ are stored in 16 bits using signed Q1.15 format or in short Q15 format. 
The generated baseband signals are normalized such that their values lie between [-1,1) and then converted to fixed-point using Q15 format. The relation between $x[k]$ (floating-point) and $x^{\prime}[k]$ (fixed-point) is given by 

\begin{equation}
    x^{\prime}[k] = \left\lfloor  {x[k]\times 2^{15}} \right\rfloor, ~ x[k] = \frac{x^{\prime}[k]}{2^{15}},
    \label{eq:float2fixed}
\end{equation}
where, $\lfloor.\rfloor$ represent the floor operation and $x^\prime[k]$'s are stored in 2's complement form.

The signal $x^{\prime}[k]$ is then rescaled before sending it to the IFFT block to get the time-domain samples. The need for rescaling stems from two factors a) DAC/ADC resolution of the RF fronted module b) the dynamic range of the IFFT block. Different RF front end modules such as USRP and various ORAN 7.2 split radio units (VVDN, LiteON, Benetel) are supported by the OAI gNB.
The fixed-point baseband signal $x^{\prime}[k]$ is scaled as
\begin{equation}
    \bar{x}[k] = \left\lfloor \frac{\text{A}\times x^{\prime}[k]}{2^{15}} \right\rfloor,
\end{equation}
where, A is a design parameter based on decibels relative to full scale (dBFS).

\subsection{Decibels relative to Full Scale}
Decibels relative to Full Scale (dBFS) is a unit of measure for the amplitude levels in digital systems \cite{dbfs}. In a Q15 fixed-point format,
the maximum represented level ${\text{A}_{\text{max}}} = 2^{15}$. To convert a level A into dBFS scale, we use the following formula

\begin{equation}
   \text{A}_{\text{dBFS}} = 20\log_{10}\left(\frac{\text{A}}{\text{A}_{\text{max}}}\right).
\end{equation}

The choice of $\text{A}_{\text{dBFS}}$ depends on the device and signal characteristics and is configurable using the parameter {\tt{tx\_amp\_backoff\_dB}} in the configuration file in OAI. The value of $\text{A}$ and $\text{A}_{\text{dBFS}}$ for a USRP and an O-RAN 7.2 VVDN RU in OAI are mentioned in Table~\ref{tab:dbfs}.

\begin{table}[htbp]
\caption{Device specific IQ bit representation in OAI}
\centering
\begin{tabular*}{\columnwidth}{@{\extracolsep{\fill}}cccc}
\toprule
Device & $\text{A}_\text{{dBFS}}$ & A & Bits\\
\midrule
USRP B210 & -36 & 519 & 9\\
O-RAN 7.2 split VVDN RU & -12 & 8231 & 13\\
\bottomrule
\end{tabular*}
\label{tab:dbfs}
\end{table}

In the case of USRP B210, the choice of $\text{A}_{\text{dBFS}}$ arises from two factors: 1. 12-bit ADC/DAC in USRP and 2. Input to the IFFT block is scaled down by 3 bits to guard against possible signal saturation during IFFT. On the other hand, the RU manufacturer mentions the $\text{A}_{\text{dBFS}}$ in the device specifications. 

We now shed light on estimating the transmit, receive powers per RE in digital and analog domains using the fixed-point tools. This is essential in calculating the link-budget, signal strength and signal-to-noise ratio (SNR) in OAI-based positioning experiments.  

\subsection{Transmit and Receive Power}
Let the reference signal transmitted be denoted by
$x^{\prime}[k],k\in\mathcal{S}$, where $\mathcal{S}$ is the set of REs where the signal is mapped, and $|\mathcal{S}|=N$.
The average power per RE is calculated as
\begin{equation}
    P_t = \frac{1}{N} \sum_{k\in\mathcal{S}}|x^{\prime}[k]|^2,
\end{equation}
The power is then converted to dBm using 
\begin{equation}
    P_{t,\text{dBm}} = 10\log_{10}(P_t) - 10\log_{10}((2^{15})^2) + 30 + G_t +G^c_t,
\end{equation}
where, the term $(2^{15})^2$ arises from the conversion from Q15, 30 appears due to the conversion of dBw to dBm, $G_t$ and $G^c_t$ are the transmit gain and calibration offset of a device respectively. 

The received power per RE in Q15 can be estimated from the received signal as
\begin{equation}\label{eq:pr}
    P_r = \frac{1}{N}\sum_{k\in\mathcal{S}}|y^{\prime}[k]|^2.
\end{equation}

Further, $P_r$ in dBm at the receiver antenna port can be obtained by,
\begin{equation}\label{eq:pr_dbm}
    P_{r,\text{dBm}} = 10\log_{10}(P_r) - 10\log_{10}((2^{15})^2) + 30 - G_r + G^c_r
\end{equation}
where, $G_r$ is the receive gain and $G^c_r$ is the calibration offset. 
Note that the calibration values $G_t^c,G_r^c$ are obtained by varying the gains $G_t, G_r$ and measuring with a spectrum analyzer, and they vary from device to device.



\subsection{SNR Estimation}
In the computation of received SNR in an OFDM system, the received signal power $P_r$, a combination of both signal and noise power per RE is obtained from \eqref{eq:pr}.
The received $\text{SNR}$ on a RE is obtained by,
\begin{equation}
    \text{SNR} = \frac{P_r-P_n}{P_n}.
\end{equation}

The noise power $P_n$ is estimated similar to \eqref{eq:pr} from empty REs where no signal is present.



\begin{figure*}
\begin{lstlisting}[language=c, caption=Code snippet of a macro definition in T\_messages.txt.,label=fig:code_snippet_t_mess]
ID = GNB_PHY_UL_FREQ_CHANNEL_ESTIMATE
    DESC = gNodeB channel estimation in the frequency domain
    GROUP = ALL:PHY:GRAPHIC:HEAVY:GNB
    FORMAT = int,gNB_ID : int,rnti : int,frame : buffer,chest_f
\end{lstlisting}
\end{figure*}

\begin{figure*}
\begin{lstlisting}[language=C, caption=T tracer based data extraction code snippet of SRS frequency domain channel estimation.,label=fig:code_snippet]
T(T_GNB_PHY_UL_FREQ_CHANNEL_ESTIMATE,
  T_INT(srs_pdu->rnti),
  T_INTframe_rx),
  T_BUFFER(srs_estimated_channel_freq[0][0], frame_parms->ofdm_symbol_size * sizeof(int32_t)));
\end{lstlisting}
\end{figure*}
\section{Data Extraction}\label{sec:data_extraction}
This section presents the data collection procedure using the {\tt{T\_tracer}} tool in OAI. It provides an example of extracting the SRS channel estimates using the tool, but the procedure is similar for any signal.

The SRS frequency domain channel estimates of length $K$, estimated using least squares are stored in a variable named {\tt{srs\_estimated\_channel\_freq[][][]}} in the function {\tt{nr\_srs\_channel\_estimation()}}. The variable is a three dimensional array, with rx antenna index as a first dimension, tx antenna index as a second dimension and subcarrier index in the third dimension. Using the {\tt{T\_tracer}} tool, data from any variable can be stored over time in a file without effecting the real-time performance of the OAI gNB and UE. The use of {\tt{T\_tracer}} for data collection is as follows,
\begin{itemize}
    \item Define an {\tt{ID}} in the {\tt{T\_messages.txt}} file, for example {\tt{GNB\_PHY\_UL\_FREQ\_CHANNEL\_ESTIMATE}} as shown in Listing~\ref{fig:code_snippet_t_mess}.
    \item Use the {\tt{ID}} in the code as {\tt{T\_ID}} and fill the function with appropriate variables as shown in Listing~\ref{fig:code_snippet}.
    \item Compile the {\tt{T\_tracer}} and gNB as described in \cite{t_tracer}.
    \item Run gNB using an additional argument {\tt{{-}{-}T\_stdout 2}}.
    \item Parallel to running the gNB, in an another terminal, run {\tt{record}} executable to record the data using the {\tt{ID}} as follows,\\
    {\tt{./record -d ../T\_messages.txt -o channel\_frequency.raw -on GNB\_PHY\_UL\_FREQ\_CHANNEL\_ESTIMATE}}
    \item After recording the data, the channel estimates from the file {\tt{channel\_frequency.raw}} can be extracted using the variable {\tt{chest\_f}} as defined in Listing~\ref{fig:code_snippet_t_mess} and {\tt{extract}} executable as follows,
    
    {\tt{./extract -d ../T\_messages.txt channel\_frequency.raw GNB\_PHY\_UL\_FREQ\_CHANNEL\_ESTIMATE chest\_f -o srschF.raw}}
\end{itemize}

Now, the frequency domain SRS channel estimates will be stored in a binary file: {\tt{srschF.raw}}. This binary file can be imported to MATLAB/OCTAVE for further offline analysis. The details of compiling {\tt{T\_tracer}} are available in \cite{t_tracer_basic}. A sample MATLAB/OCTAVE script that reads the binary file and plots the SRS channel estimates is provided in \cite{rtt_dataset}. More details on the usage of {\tt{T\_tracer}} can be found in \cite{t_tracer}.

We now present a use-case in terms of an RTT-based range estimation prototype with OAI \cite{Mundlamuri2024} leveraging the tools and details outlined before.

\begin{figure}[t]
\centerline{\includegraphics[width=3.5in]{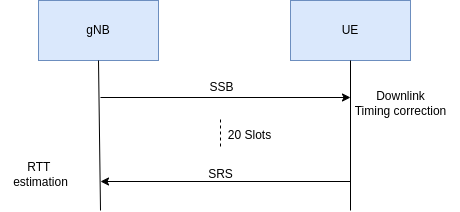}}
\caption{RTT implementation in OAI phy-test mode.}
\label{fig:phy_test_oai}
\end{figure}

\begin{figure*}
    \centering    \includegraphics[width=\textwidth]{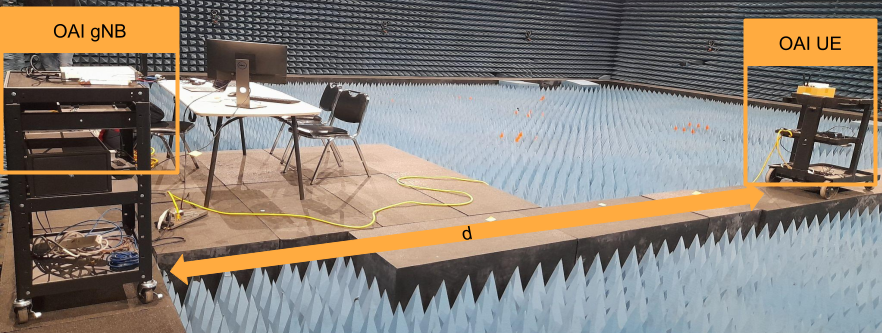}
    \caption{Experimental setup in the anechoic chamber.}
    \label{fig:chamber}
\end{figure*}

\section{RTT Estimation Prototype}\label{sec:prototype}


Recently, we have proposed a novel RTT estimation method with less overhead and latency compared to the existing schemes in the standards \cite{Mundlamuri2024}. In this section, we provide the details of the a) experimental setup, b) data extraction process and tools, c) collected dataset

The experimental setup consists of a single antenna gNB and a UE in a line-of-sight (LoS) leveraging the OAI 5G NR protocol stack \cite{KalaloAbhiLuh_20} and USRP B210 software-defined radio cards to build the gNB and UE. Furthermore, the SC2430 NR signal conditioning module is employed as an external RF front-end at the gNB \cite{sc_mod}. The experimental prototype incorporates a novel RTT procedure introduced in \cite{Mundlamuri2024}, designed to provide good performance in low SNR scenarios by coherently combining multiple measurements. This procedure involves the UE synchronizing with the downlink synchronization signal block (SSB) and transmitting SRS, as depicted in Figure~\ref{fig:phy_test_oai}. More details about the RTT estimation scheme can be found in \cite{Mundlamuri2024}.

The experiments were carried out in an anechoic chamber, as depicted in Figure~\ref{fig:chamber}, using the 5G NR system parameters shown in Table~\ref{tab:1}. During the experiment, the gNB remained stationary while the UE was moved in 1-meter increments, as shown in Figure~\ref{fig:chamber}. Variation in uplink signal-to-noise ratio (SNR) was achieved by adjusting the USRP transmit (TX) gain. Multiple SRS channel estimates were obtained at each distance, and the data was stored for further offline analysis. We have used the {\tt{T\_tracer}} tool to extract the required data from OAI.

\begin{table}[htbp]
\caption{System Parameters}
\centering
\begin{tabular*}{\columnwidth}{@{\extracolsep{\fill}}cc}
\toprule
Parameters & Values \\
\midrule
System bandwidth & 38.16 MHz \\
Subcarrier Spacing & 30 KHz \\
Centre frequency & 3.69 GHz \\
Sampling rate($f_s$) & 46.08 MHz \\
FFT size ($K$) & 1536 \\
Cyclic prefix & 132\\
SSB bandwidth & 7.2 MHz\\
SRS bandwidth & 37.44 MHz\\
SRS comb size & 2\\
\bottomrule
\end{tabular*}
\label{tab:1}
\end{table}

\subsection{Dataset}\label{sec:dataset}
\begin{figure}[t]
\centerline{\includegraphics[width=3.5in]{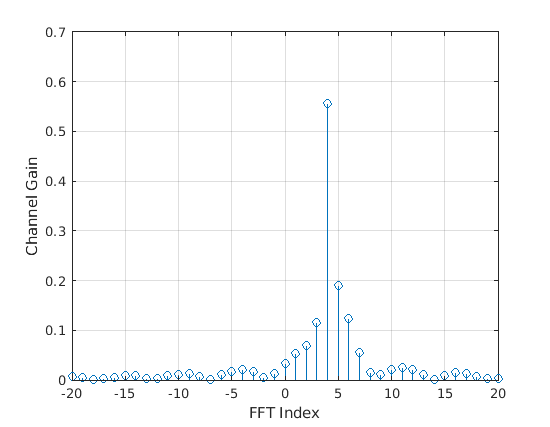}}
\caption{Impulse response}
\label{fig:ch_est}
\end{figure}

The dataset collected using the RTT prototype mentioned above in an anechoic chamber is made public in \cite{rtt_dataset}. The dataset includes multiple measurements at different distances varying from 7-11m in 1 m increments. At each distance, SNR is varied by changing the TX gain from 39.5 dB to 89.5 dB of the USRP in steps of 10 dB. The SNR corresponds to 89.5 is 25 dB. In the published dataset\cite{rtt_dataset}, the folder name contains the corresponding distance between gNB, UE, and the TX gain used. The TX gain can be inferred from the folder name as follows: the sub-string {\tt{ue\_att\_x}} should be interpreted as x dB attenuation from the maximum gain of the USRP B210, which is 89.5 dB. For example, {\tt{ue\_att\_0}} corresponds to the TX gain 89.5 dB, and {\tt{ue\_att\_50}} corresponds to the TX gain 39.5 dB. Each folder contains the following recorded data files in Q15 format,
\begin{itemize}
    \item {\tt{srs\_chF.raw}}: Frequency domain least square channel estimates of the SRS in comb fashion.
    \item {\tt{srs\_chF\_lin\_interp.raw}}: Frequency domain least square channel estimates of the SRS in comb fashion and the channel estimates of the subcarriers between the two SRS symbols in a comb are linearly interpolated.
    \item {\tt{srs\_chT.raw}}: Impulse response of the SRS.
    \item {\tt{noise.raw}}: Noise measured in an empty OFDM symbol.
\end{itemize}
To read and analyze these files, a sample MATLAB script is provided in \cite{rtt_dataset}. For example, an impulse response sample from the dataset at 10m and a USRP TX gain of 89.5 dB is shown in Figure~\ref{fig:ch_est}. The x-axis in Figure~\ref{fig:ch_est} represents the IFFT index in samples. The RTT in samples between the gNB and UE can be estimated from the peak $p$ of the impulse response. Finally, the range estimate $\hat{d}$ between the gNB and UE from peak $p$ is obtained by,
\begin{equation}
    \hat{d} = \frac{p\times c}{2f_s}
\end{equation}

where $c$ is the speed of light and $f_s$ is the sampling rate. Utilizing the system configuration mentioned in Table~\ref{tab:1}, the estimated range is 13.02m. 

Note that in this example, the estimated range is based on the maximum peak of the channel impulse response from a single measurement. However, better accuracy in terms of range estimation error can be achieved by obtaining multiple measurements using the signaling mechanism described in Figure~\ref{fig:phy_test_oai}. The range estimation combining multiple measurements ($M$) coherently using peak detector (PD) and matched filter (MF)\cite{Mundlamuri2024} techniques are described as follows,
\begin{equation}
\hat{d}\ (PD)  = \frac{\text{c}}{2f_sM}\sum_{m=1}^{M} \argmax \left| IDFT \left\{\bm{\hat{h}}_m\right\}\right|,
\end{equation}
and
\begin{equation}
    \hat{d}\ (MF) = \frac{c}{2f_s}\times\argmax_{{\tau}} \frac{1}{{M}}\sum^{{M}}_{m=1}|\bm{v}(\tau)^{\He}\bm{\hat{h}}_m|^2,
\end{equation}
where, $\bm{\hat{h}}_m,m\in[1,M]$ is the $m$-th SRS channel frequency response measurement, $f_s$ is the sampling rate, $IDFT\{.\}$ denotes the Inverse Discrete Fourier Transform, $\bm{v}(\tau)=
[1,e^{-j2\pi \Delta f \tau},\ldots,e^{-j2\pi (K-1)\Delta f \tau}]^\Tr$ and $c$ is the speed of light.

\begin{figure}[t]
\centerline{\includegraphics[width=3.5in]{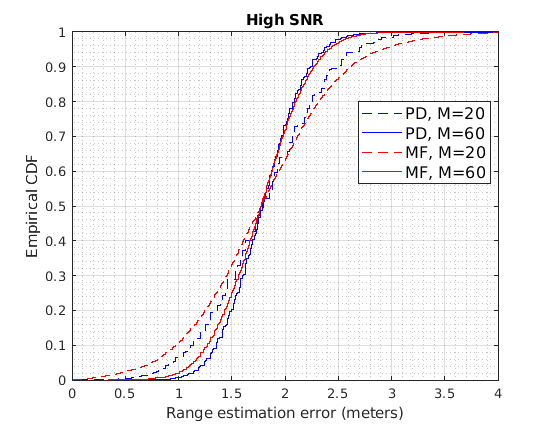}}
\caption{CDF of the range
estimation error.}
\label{fig:high_snr}
\end{figure}

\begin{figure}[t]
\centerline{\includegraphics[width=3.5in]{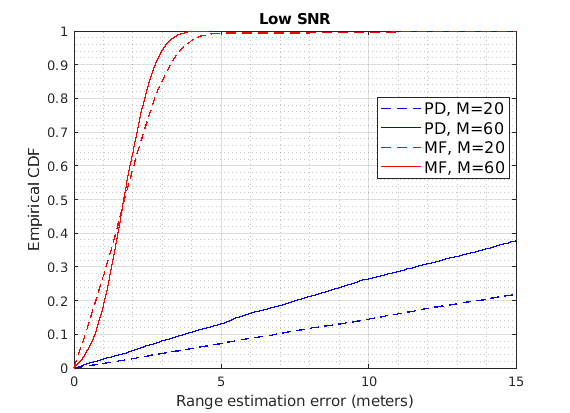}}
\caption{CDF of the range
estimation error.}
\label{fig:low_snr}
\end{figure}

The performance of the collected data can be assessed using empirical cumulative distribution function (CDF) plots of the range estimation error. The CDF plots of the range estimation error with $M=20$ and $M=60$ at high and low SNR scenarios can be seen in Figure~\ref{fig:high_snr} and Figure~\ref{fig:low_snr}, respectively. Here, the high SNR scenario refers to the USRP TX gain of 89.5 dB, where the SNR is estimated to be 25 dB, and the low SNR scenario refers to the USRP TX gain of 39.5 dB. From Figure~\ref{fig:low_snr}, It is evident that the MF outperforms the PD significantly in low SNR scenarios when multiple measurements are coherently combined.

\section{Conclusion}\label{sec:conclusion}
In our work, we delved into the essential functions related to positioning in OAI. We provided a detailed description of a baseband signal representation in OAI using fixed-point notation and demonstrated the conversion process between floating and fixed-point notation. Additionally, we offered detailed information on physical layer metrics such as TX and RX power estimation and SNR estimation in OAI. We also presented the {\tt{T\_tracer}} tool for data extraction in OAI. Furthermore, we showcased a prototype that implements the proposed novel RTT estimation mechanism using OAI. 
Lastly, we made the dataset publicly available.
\section*{Acknowledgements}
This work is partially funded by the 5G-OPERA project through the German Federal Ministry of Economic Affairs and Climate Action (BMWK) as well as the French government as part of the ``France 2030" investment program. This work is also partially supported by OUSD (R\&E) through Army Research Laboratory Cooperative Agreement Number W911NF-24-2-0065. The views and conclusions contained in this document are those of the authors and should not be interpreted as representing the official policies, either expressed or implied, of the Army Research Laboratory or the U.S. Government. The U.S. Government is authorized to reproduce and distribute reprints for Government purposes notwithstanding any copyright notation herein.
\bibliographystyle{IEEEtran}
\bibliography{References}

\end{document}